\documentclass[prl,two column,showpacs,preprintnumbers,amsmath,amssymb,revtex4-1]{revtex4-1}
\usepackage{amsfonts, amsmath, bm, bbm, graphicx, epsfig, epstopdf}
\usepackage{dcolumn}
\usepackage{textcomp}
 \usepackage[caption=false]{subfig}
\usepackage{mathtools}
\usepackage{multirow}   

\usepackage{xcolor}
    \usepackage{braket}
\renewcommand\bra[1]{{\langle{#1}|}}
\makeatletter
\renewcommand\ket[1]{%
  \@ifnextchar\bra{\k@t{#1}\!}{\k@t{#1}}%
}
\newcommand\k@t[1]{{|{#1}\rangle}}
\makeatother

\begin{document}
\title{Quasi-Contact Forces with Resonant Range Control in Rydberg Atoms}

\author{Mohammadsadegh Khazali}
\affiliation{Department of Physics, University of Tehran, Tehran 14395-547, Iran}
\affiliation{Email: mskhazali@ut.ac.ir}

\begin{abstract}
This article introduces a novel method to engineer sharply peaked, distance-selective interactions between neutral atoms by exploiting interaction-induced resonances within a resonantly driven Rydberg ladder system. By tuning laser parameters, a subsystem eigenstate twist rapidly and brought into degeneracy with the atomic ground state at precisely defined interatomic separations, resulting in an effective potential sharply localized around this resonance distance. Unlike previous off-resonant macrodimer-based schemes, our approach significantly enhances interaction strength, reaching MHz scales, and provides straightforward experimental tunability without requiring sub-wavelength positional control. Analytic expressions, validated through comprehensive master-equation simulations, detail the interaction profile's amplitude, width, and resonant distance. This precise control facilitates parallel entangling gates crucial for measurement-based quantum computing and enables simulation of complex lattice Hamiltonians with customizable connectivity. 
\end{abstract}

\maketitle

 {\bf Introduction}-- 
 Highly excited Rydberg atoms offer precisely controllable interactions that have substantial implications for advancing quantum technology. These interactions are inherently long-range, which can lead to problematic cross-talk between nonadjacent qubits, adversely affecting the fidelity of parallel quantum gates and complicating the implementation of quantum lattice models with customized connectivity. Thus, there is significant interest in engineering interaction potentials that are sharply peaked at a specific, chosen interatomic distance, effectively suppressing unwanted long-range interactions.
Such distance-selective interactions exhibit a spatial profile reminiscent of a delta function, sharply peaked at a chosen interatomic separation with negligible coupling elsewhere. Implementing delta-function-like potentials not only minimizes detrimental long-range tails but also ensures precise interactions confined to designated neighbors. 

This property is particularly advantageous for performing parallel multi-qubit gates, especially those required for Measurement-based quantum computing (MBQC), which necessitates the rapid generation of large-scale, high-fidelity entanglement among a register of long-lived qubits. The distance-selective interaction allows parallel entangling operation with global pulses in the atomic lattice. The proposed interaction enables geometric entanglement filtering, expediting and parallelizing the syndrome extraction in surface error correction codes within a geometrically designed atomic lattice with global laser addressing.
 The proposed interaction enables the simulation of intricate lattice models, such as extended Hubbard, Ising, or Su–Schrieffer–Heeger (SSH) Hamiltonians, with customizable interaction graphs \cite{Kha22,Hun16,Che20}. Consequently, such programmable interaction networks significantly enrich the exploration of quantum phase transitions \cite{Bai16}, topological states of matter \cite{Kha22,Hun16}, quantum annealing processes \cite{Lu25}, simulation of photosynthesis light harvesting \cite{Moh08}, and Hopfield networks.

\begin{figure*}
\centering
    \includegraphics[trim=20 300 190 40, clip, width=\linewidth]{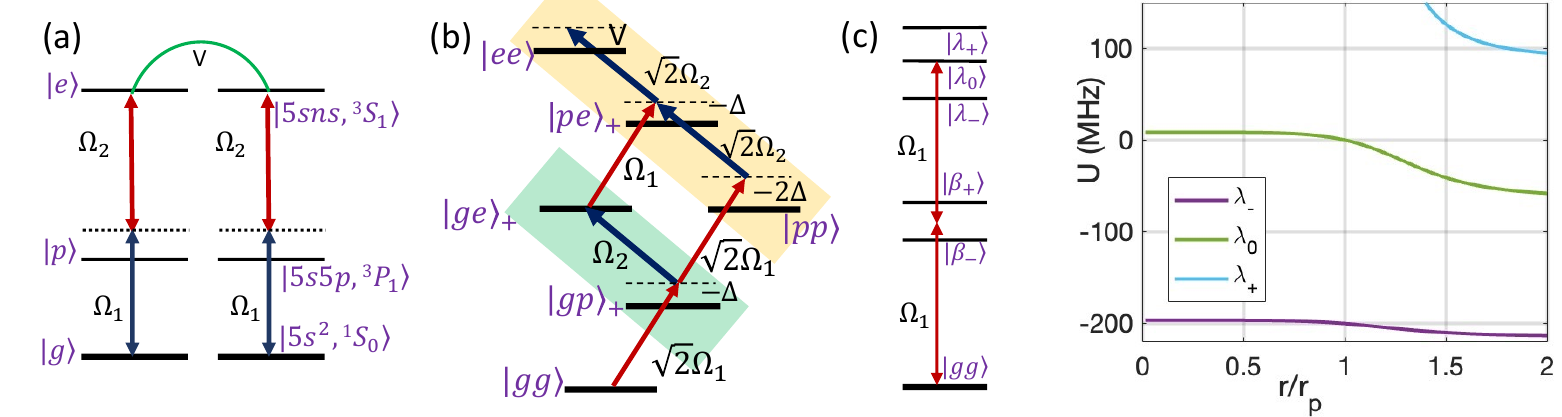}
\caption{ {\bf Level scheme for generating distance-selective, delta-function-like interactions.}
(a) Two-photon resonant excitation of Rydberg atoms with strong interactions enables the engineering of tailored interaction profiles. Notably, in the regimes $\Omega_2 / 2\Delta > 1$ for positive detuning and $\Omega_2 / 2|\Delta| < 1$ for negative detuning, the resulting interaction acquires a sharply peaked, delta-function-like shape (see Fig.~2a).
(b) This sharply localized interaction peak originates from an interaction-induced resonance. When $\Omega_1 \ll \Omega_2$, the two-atom Hilbert space naturally separates into three subspaces, i.e., the ground state, singly excited states (green boxes), and doubly excited states  (yellow box), with transitions mediated by the weak $\Omega_1$ field.
(c) The interplay between strong coupling $\Omega_2$ and interatomic interaction $V$ becomes evident upon diagonalizing the singly and doubly excited subspaces. This yields the dressed eigenstates $|\beta_{\pm}\rangle$ and $|\lambda_{0,\pm}\rangle$, respectively. (d) The interaction shifts the $|\lambda_{0}\rangle$ state into resonance with the ground state at a specific interatomic distance $r_p$, see Eq.~\ref{Eq_resonance}. This interaction-induced resonance sharply enhances the excitation probability (Fig.~2b), effectively producing an ultra-narrow, distance-selective interaction peak resembling a delta function (Fig.~2a).
}\label{Fig1}
\end{figure*}

In the continuum limit, distance-selective interactions are analogous to zero-range potentials positioned at finite radii, known as {\it shell potentials}. Such $\delta$-shell potentials exhibit distinctive scattering properties and are featured in exactly solvable scattering theory models \cite{Erm22,Sto05,Wit05}. Additionally, off-centered contact interactions have been theoretically explored for novel phenomena, such as controlled particle bunching and anti-bunching in a Harmonic trap \cite{Bou24}, indicating promising opportunities for investigating unconventional scattering and bound-state physics at larger separations.

In this work, we introduce a novel method to realize distance-selective interactions by exploiting the fast twist of subsystem eigenstates in a resonantly driven Rydberg ladder configuration. Specifically, interaction-induced resonances form degeneracy between a manifold of doubly excited states and the ground state at precisely controlled interatomic distances, generating an effective potential sharply peaked at a target separation. Importantly, the two-Rydberg excitation in this manifold is absent, ensuring the perseverence of trapping. Furthermore, we study three-body resonance peaks illuminating the technique that could be extended to higher-order many-body resonances. In particular, such genuine many-body resonances enable single-step, parallel stabilizer operations \cite{Kha24Terminal}, direct implementation of multi-qubit gates \cite{Kha20}, and the exploration of exotic many-body dynamics that cannot be reached through pairwise interactions alone. We provide analytic expressions complemented by numerical simulations, thoroughly characterizing the interaction profile’s amplitude, width, and resonant separation, which are all controllable by driving laser parameters.

Unlike previously studied soft-core potentials obtained through resonant Rydberg dressing \cite{Gau16, Kha21, Shi24}, the current scheme operates in a distinct parameter regime to produce sharply localized, off-centered $\delta$-function-like potentials, see discussion. 
Previous approach for making distance-selective interaction, applied off-resonant dressing of ground-state atoms into Rydberg macrodimer states \cite{Hol22}, faced limitations including atom loss, continuum coupling, and demanding experimental controls. In contrast, our resonant scheme achieves orders-of-magnitude improvements in interaction sharpness and strength, simplifies experimental implementation, and enables precise, rapid tuning of the interaction profile without requiring stringent atomic position or orientation control, see discussion.

 {\bf Scheme}--
We consider two strontium atoms, each driven by the resonant two-photon ladder scheme of Fig.~1a to the Rydberg state $\ket{e}$ via the intermediate state $\ket{p}$.  In the rotating frame, the single-atom Hamiltonian reads
\begin{equation}
H_i = \frac{\Omega_1}{2}\bigl(\sigma^i_{gp} + \sigma^i_{pg}\bigr) +\frac{\Omega_2}{2}\bigl(\sigma^i_{pe} + \sigma^i_{ep}\bigr)
-\Delta\,\sigma^i_{pp},
\label{Eq_Hi}
\end{equation}
with $\sigma_{\alpha\beta}\equiv\ket{\alpha}\bra{\beta}$, Rabi frequencies $\Omega_1$ (ground–$\ket{p}$) and $\Omega_2$ ($\ket{p}$–$\ket{e}$), and detuning $\Delta$ from the intermediate level.  In the idealized, lossless limit, the system adiabatically follows the dark eigenstate
$\ket{d}\propto\Omega_2\ket{g}-\Omega_1\ket{e}$,  
which acquires no light shift.  For $\Omega_1\ll\Omega_2$, atoms remain predominantly in $\ket{g}$ but are weakly dressed by the Rydberg state, with excited–state admixture
$\,P_e\simeq(\Omega_1/\Omega_2)^2$.
When two atoms are both in $\ket{e}$, they interact via the van der Waals potential
$V_{ij} \;=\; \frac{C_6}{r_{ij}^6}\,\sigma_{ee}^i\,\sigma_{ee}^j$
where $r_{ij}$ is their separation.
To capture realistic decoherence we include spontaneous‐emission channels as Lindblad operators
$L_p = \sqrt{\gamma}\,\ket{g}\bra{p}\,, 
L_r = \sqrt{\gamma_r}\,\ket{p}\bra{e},$
with $\gamma/2\pi=7.6$ kHz for the $\ket{p}\to\ket{g}$ decay in strontium, and $\gamma_r$ for the Rydberg $\ket{e}\to\ket{p}$ decay taken from \cite{Kun93}.

We evaluate the steady‐state two‐atom density matrix $\rho_{ij}$ under continuous driving and define the total light‐shifted energy
\begin{equation}
U(r_{ij}) = Tr \bigl [\rho_{ij}(H_i + H_j + V_{ij})\bigr],  \label{Eq_Unum}
\end{equation}
where $H_{i,j}$ are the single‐atom Hamiltonians of Eq.~\ref{Eq_Hi} in the two‐photon ladder scheme of Fig.~1(a) and $V_{ij}$ is the van‐der Waals interaction between Rydberg states.

In the regime $\Omega_1\ll\Omega_2$, the two‐atom Hilbert space splits into three manifolds weakly coupled by $\Omega_1$. These manifolds are the ground $\{\ket{gg}\}$, the single‐excitation manifold spanned by $\ket{gp}_+\!=\!(\ket{gp}+\ket{pg})/\sqrt2$ and $\ket{ep}_+$ (highlighted by green in Fig.~1b), and the double‐excitation manifold $\{\ket{pp},\,\ket{pe}_+,\,\ket{ee}\}$ (highlighted by yellow).  
The strong coupling $\Omega_2$, mixes the states in each subspace. 
Pre-diagonalizing the subsystems quantifies the light-shifts experienced by the eigen-states at different interatomic separations, see Fig.~\ref{Fig1}c.
 In the single‐excitation block, the coupling Hamiltonian  in \{$\ket{gp}_+$, $\ket{ge}_+$\} basis
has eigenvalues $\beta_\pm/\hbar=-\tfrac\Delta2\pm\tfrac12\sqrt{\Delta^2+\Omega_2^2}$, which remain off‐resonant with $\ket{gg}$.

\begin{figure*}
\centering
    \includegraphics[trim=0 360 395 0, clip, width=\linewidth]{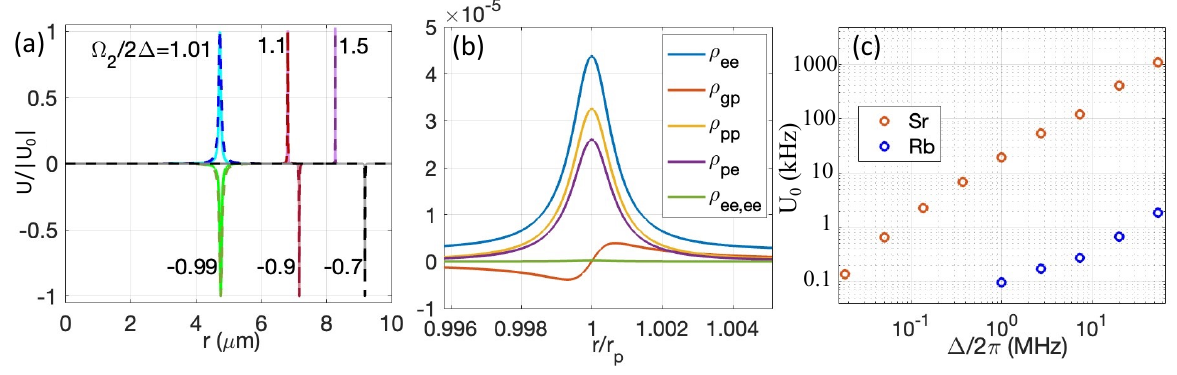}
\caption{ {\bf Interaction profile} (a) The analytic (dashed line) and numeric (solid line) interaction profile are presented based on Eq.~\ref{Eq_Lorentzian} and \ref{Eq_Unum}, respectively. With $\Omega_1$ and $\Delta$ held constant, varying $\Omega_2$ tunes the amplitude, sign, and resonant separation $r_p$ of the effective interaction $U(r)$. As explained in Eq.~\ref{Eq_w}, deviation of $\Omega_2/2\Delta$ from 1 makes the spatial peak profile sharper and increases the contact interaction range $r_p$. This tunability allows designing different interaction graphs in a lattice.  (b) Distance‐dependent steady‐state populations (density‐matrix elements) as a function of interatomic spacing $r$.  The Lorentzian peak emerges from the interaction‐induced two‐body resonance in the doubly‐excited manifold (yellow subspace in Fig.~1b), which sharply enhances the intermediate state population and hence the system energy at $r=r_p$. 
(c) Maximum interaction strength at a decoherence rate of $100s^{-1}$ achievable for Sr (upper circles) and Rb (lower circles). Here $\Omega_1$ is tuned to adjust $100s^{-1}$ decoherence rate at the peak position. For Struntium, the applied parameter is in the 35kHz$<\Omega_1/2\pi<$210kHz range. Interaction to loss ratio scales by $\Delta/\gamma$.
Applied Parameters in (a,b): $\Delta/2\pi=10$ MHz, $\Omega_1=200$ kHz, principal quantum number $n=100$; panel (b,c) uses $\Omega_2/(2\Delta)=1.1$.
}\label{Fig2}
\end{figure*}

By contrast, in the yellow shaded manifold of double‐excitation subspace, the $3\times3$ Hamiltonian in \{$\ket{pp}$, $\ket{pe}_+$, $\ket{ee}$\} basis
\begin{equation}
\frac{H_{\text{yellow}}}{\hbar}=\begin{pmatrix}
-2\Delta-i\gamma & \sqrt{2}\Omega_2/2 &0\\
 \sqrt{2}\Omega_2/2& -\Delta-i\gamma/2 & \sqrt{2} \Omega_2/2 \\
 0&\sqrt{2}\Omega_2/2  & V(r) \\
\end{pmatrix}
\label{Eq_S3}
\end{equation}
supports $\ket{\lambda_{\pm,0}}$  eigenstates. 
Considering the space-dependent interaction $V=C_6/r^6$ and in the regime of $\Delta\gg\gamma$, Fig.~1d shows how $\lambda_0$ eigenvalue crosses zero at the resonance condition
\begin{equation}
V(\mathrm{r_{p}})=\frac{2\Delta \Omega_2^2}{(\Omega_2^2-4\Delta^2+\gamma^2)}.
\label{Eq_resonance}
\end{equation}
At this separation $r=r_p$, the corresponding eigenstate
\begin{equation}
\ket{\lambda_0(r_p)} =
\frac{\Omega}{\sqrt{8}\Delta}|pp\rangle
+|pe\rangle_{+}-\frac{\Omega^2  - 4\Delta^2+ \gamma^2}
{\sqrt{2}\Omega\, 2\Delta }|ee\rangle
\label{Eq_EigSate}
\end{equation}
becomes degenerate with $\ket{gg}$, producing an avoided crossing that manifests as a sharply peaked, delta‐function light shift in $U(r)$. This interaction‐induced resonance thus generates a distance‐selective potential whose width and amplitude are controlled by $\Delta$ and $\Omega_2$. Fig.~\ref{Fig2}b represents the evolution of single atom density matrix elements at the position of resonance. Clearly, at the position of resonance, the intermediate state gets populated. In case that $\Omega_2^2-4\Delta^2\ll \Omega_2^2$, then the double Rydberg‐excitation probability from Eq.~\ref{Eq_EigSate} remains small, which is essential for preserving the trap. In this regime, the relevant figure of merit i.e., the ratio of coherent interaction strength to loss rate, is simply $\Delta/\gamma$.

The solitary sharp peak happens in two regimes of parameters ($\Omega_2>2\Delta$ with $\Delta>0$) and ($\Omega_2<2|\Delta|$ with $\Delta<0$), see Fig.~\ref{Fig2}a.
It forms a Lorentzian peak
\begin{equation}
\label{Eq_Lorentzian}
U=\frac{U_0}{1+(r-r_p)^2/w^2}.
\end{equation}
As the peak is formed by populating the intermediate state, the width of this resonant peak $w$ is defined by the line width of the intermediate state. Considering the interatomic deviation of $w$ from the resonance point $r_p$, the corresponding deviation of the van der-Waals interaction  $\delta V=C_6/r_p^6-C_6/(r_p\pm w)^6$  causes a level shift of $ \lambda_0(r_p\pm w)=\gamma_p$, which breaks the resonance. This displacement $w$ defines the HWHM of the Lorentzian interaction $U$.
If the interaction perturbs by a small $\delta V$ around $V(r_p)$, $\lambda_{0}$ grows linearly with $\delta V$ as
\begin{equation}
\lambda_{0}(V(r_p)+\delta V)\approx \frac{(\Omega_{2}^{2}-4\Delta^{2})^{2}}{2\,(\Omega_{2}^{4}+8\Delta^{4})}\;\delta V.
\end{equation}
Considering Eq.~\ref{Eq_resonance}, and $w\ll r_p$, the ratio of the Lorentzian width to the resonant inter-atomic separation would be given by 
\begin{equation}
(w/r_p)\approx \frac{ \gamma (\Omega_{2}^{4}+8\Delta^{4})}{6\Delta\Omega_{2}^{2}(\Omega_{2}^{2}-4\Delta^{2}+\gamma^2)}.
\label{Eq_w}
\end{equation}
Hence, deviating from $\Omega_2=2\Delta$ reduces the $w/r_p$ and makes the interaction more $\delta$-function-like, see Fig.~\ref{Fig2}a for examples.
Selecting a long-lived clock state, with a correspondingly small decay rate $\gamma$ as the intermediate level, yields a much sharper resonance peak.
The depth of interaction at the peak position could be obtained perturbatively $(\epsilon=\Omega_1/\Omega_2\ll1$), as
\begin{eqnarray}
U_0=\frac{3\,\Delta\,\Omega_{1}^{4}\,\bigl[\;4\,\Delta^{2}\,(4\,\Delta^{2} + \gamma^{2} + 3\,\Omega_{2}^{2}) + \gamma^{2}\,\Omega_{2}^{2}\bigr]}
{2\,\gamma^{2}\,\Omega_{2}^{2}\,\bigl[\,4\,\Delta^{2}\,\gamma^{2} + (\,4\,\Delta^{2} + \Omega_{2}^{2}\,)^{2}\bigr]}.
\label{Eq_U0}
\end{eqnarray}
Equations \ref{Eq_resonance}, \ref{Eq_Lorentzian}, \ref{Eq_w}, and \ref{Eq_U0} together capture both the line shape and the maximum interaction strength in these two detuning regimes.

\begin{figure}
\centering
    \includegraphics[trim=0 270 494 0, clip, width=\linewidth]{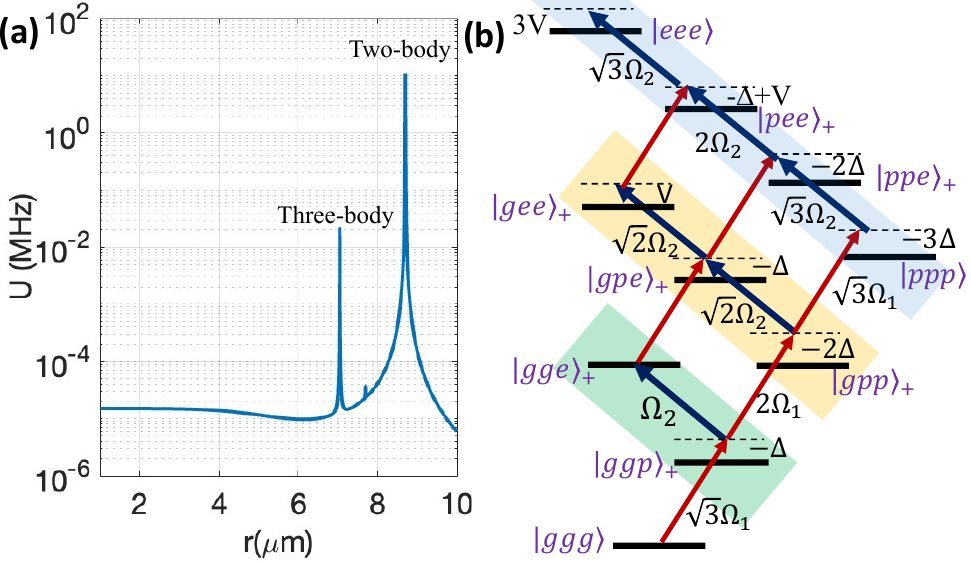}
\caption{ {\bf Three-Body Interaction} (a) Three- and two-body resonance peaks in the interaction profile of three atoms in an equilateral triangle geometry. (b) Level scheme in a three-atom collective basis. Here the Rydberg state n=100 is targeted by the driving parameters $\Omega_1/2\pi=318$kHz, $\Delta/2\pi=10$MHz, $\Omega_2=4\Delta$.}\label{Fig3}
\end{figure}

Considering three atoms in an equilateral triangle geometry, three-body interaction resonances emerge alongside the familiar two-body peaks (Fig.~\ref{Fig3}a). Under the collective-level scheme depicted in Fig.~\ref{Fig3}b, choosing $\Omega_{1}\ll \Omega_{2}$ effectively divides the system into four decoupled subspaces. In this regime, the blue manifold becomes resonant with the collective ground state at a specific interatomic distance $r_{p3}$, defined by
\begin{equation}
V(r_{p3})=
\frac{2\Delta^3 - 5\Delta\Omega^2 \pm \sqrt{4\Delta^6 - 4\Delta^4\Omega^2 + 9\Delta^2\Omega^4 + 4\Omega^6}}
{2(2\Delta^2 - \Omega^2)}.
\end{equation}
The fast twist of the resonant eigenstate in this manifold generates the three-body interaction peak.
By tuning parameters to access a specific many-body resonance, one can engineer tailored many-body connectivity within the lattice.

{\bf Discussion}--
Comparing the current approach with the previous realization of distance-selective interactions based on off-resonant coupling to Rydberg macrodimers \cite{Hol22}, several distinct advantages emerge.
The  {\it potential width} in our scheme is controlled by the laser parameters and the intermediate state linewidth. In contrast, the macrodimer-based scheme exhibits a finite-width peak, fundamentally limited by the macrodimer vibrational wavefunction and atomic motional states, thus requiring sub-wavelength control of atom positions.

Due to direct resonance-enhanced coupling at the target distance $r_p$, our method achieves substantially higher {\it potential depth}, potentially reaching the MHz scale. This enhancement represents orders-of-magnitude improvement over the macrodimer-based dressing approach, which yielded moderate couplings on the order of a few hundred hertz, inherently limited by its second-order, off-resonant nature.

Our technique offers flexible and straightforward {\it resonant distance tunability} $r_p$ via adjustments of laser detunings or the selection of different Rydberg pair resonances. We provide analytical expressions explicitly relating system parameters to $r_p$, enabling precise targeting of nearest-neighbor or next-nearest-neighbor interactions. The macro-dimer scheme, on the other hand, determines $r_p$ by the fixed macrodimer bond length of the chosen vibrational state, making adjustments more cumbersome and necessitating the selection of different molecular states or atomic species.

Our resonant ladder scheme allows {\it high-precision control} over the interaction via laser intensity and detuning, enabling rapid switching and fine-tuning of the interaction peak position, width, and strength. In comparison, the macrodimer scheme’s spatial sharpness was constrained by atomic motional spread and required intricate experimental control, including ground-state cooling, careful alignment of atomic orientations, and precise management of multiple laser frequencies and polarizations.

On a different aspect, we should compare this work with the earlier works on resonant Rydberg dressing \cite{Gau16,Kha21,Shi24}, where they achieved soft-core interaction profiles. Those softcore potentials feature flat-topped, non-divergent profiles that saturate at short range. They could be observed when $\Omega_2<2|\Delta|$ for positive detuning and $\Omega_2>2|\Delta|$ for negative detuning. They explore the case that interaction is so large that it decouples $\ket{ee}$ within the soft core. They found the maximum softcore interaction strength for a specific laser intensity of $\Omega_2=2|\Delta|$ where $\lambda_-$ in Fig.~\ref{Fig1}c gets in resonance with $\ket{gg}$. 
This resonance has been deployed for making an optical lattice using a standing wave driving field \cite{Kha25}.
The current work is distinct in generating off-centered $\delta$-like peak potential away from zero separation rather than a soft-core profile. Here we are working in a different regime with opposite detunings, i.e. $\Omega_2<2|\Delta|$ for negative detuning and $\Omega_2>2|\Delta|$ for positive detuning, the interaction profile forms a distance-selective delta function peak.
Here, we discuss that the finite and specific value of interaction makes the other eigenstate  $\lambda_0$ (see Fig.~\ref{Fig1}) in resonance with $\ket{gg}$ at a specific interatomic distance, causing a peak potential.

{\bf Outlook}--  Rydberg interactions have already proven indispensable across a wide range of quantum-technology applications \cite{KhaFast24,Kha23Lech,Kha23,Mom25,Kha21IJAP,Kha19,Kha18,Kha17,Kha16,Kha15,KhaThesis}. By introducing a tunable, sharply peaked interaction profile centered at a chosen separation, our work adds a powerful new tool to this experimental and theoretical toolbox.
In addition to the applications discussed in the Introduction, the introduced distance‐selective interaction scheme opens a new avenue for creating diatomic molecules with micrometer‐scale bond lengths \cite{Kha21}.  This offers a unique window into molecular physics, e.g., to observe vibrational motion of this giant molecule in real time, or use local fields to perturb one atom and see the effect on the bond. Such studies are normally impossible for conventional molecules since bond lengths are too small and internal dynamics are too fast.


\begin{thebibliography}{99}
\bibitem{Kha22} M.~Khazali, Discrete-time quantum-walk \& Floquet topological insulators via distance-selective Rydberg-interaction, {\it Quantum} {\bf 6}, 664 (2022).

\bibitem{Hun16} C.~L.~Hung, A.~González-Tudela, J.~I.~Cirac, and H.~J.~Kimble, Quantum spin dynamics with pairwise-tunable, long-range interactions, {\it Proc. Natl. Acad. Sci. U.S.A.} {\bf 113}, E4946 (2016).

\bibitem{Che20} H.~J.~Chen, Y.~Q.~Yu, D.~C.~Zheng, and R.~Liao, Extended Bose-Hubbard model with cavity-mediated infinite-range interactions at finite temperatures, {\it Sci. Rep.} {\bf 10}, 9076 (2020).

\bibitem{Bai16} S.~Baier {\it et al.}, Extended Bose-Hubbard models with ultracold magnetic atoms, {\it Science} {\bf 352}, 201 (2016).

\bibitem{Lu25} Y.~Lu, W.~Chen, S.~Zhang, K.~Zhang, J.~Zhang, J.~N.~Zhang, and K.~Kim, Implementing Arbitrary Ising Models with a Trapped-Ion Quantum Processor, {\it Phys. Rev. Lett.} {\bf 134}, 050602 (2025).


\bibitem{Moh08} M.~Mohseni, P.~Rebentrost, S.~Lloyd, and A.~Aspuru-Guzik, Environment-assisted quantum walks in photosynthetic energy transfer, {\it J. Chem. Phys.} {\bf 129}, 174106 (2008).

\bibitem{Erm22} F.~Erman and S.~Seymen, A direct method for the low energy scattering solution of delta shell potentials, {\it Eur. Phys. J. Plus} {\bf 137}, 330 (2022).

\bibitem{Sto05} R.~Stock, A.~Silberfarb, E.~L.~Bolda, and I.~H.~Deutsch, Generalized pseudopotentials for higher partial wave scattering, {\it Phys. Rev. Lett.} {\bf 94}, 023202 (2005).

\bibitem{Wit05} D.~Witthaut, S.~Mossmann, and H.~J.~Korsch, Bound and resonance states of the nonlinear Schrödinger equation in simple model systems, {\it J. Phys. A: Math. Gen.} {\bf 38}, 1777 (2005).

\bibitem{Bou24} G.~Bougas, N.~L.~Harshman, and P.~Schmelcher, Impact of dark states on the stationary properties of quantum particles with off-centered interactions in one dimension, {\it Phys. Rev. A} {\bf 110}, 023327 (2024).

\bibitem{Gau16} C.~Gaul, B.~J.~DeSalvo, J.~A.~Aman, F.~B.~Dunning, T.~C.~Killian, and T.~Pohl, Resonant Rydberg Dressing of Alkaline-Earth Atoms via Electromagnetically Induced Transparency, {\it Phys. Rev. Lett.} {\bf 116}, 243001 (2016).

\bibitem{Kha21} M.~Khazali, Rydberg noisy dressing and applications in making soliton molecules and droplet quasicrystals, {\it Phys. Rev. Research} {\bf 3}, L032033 (2021).

\bibitem{Shi24} Z.~Shi, M.~Khazali, L.~Qin, Y.~Zhou, and Y.~Zhong, Pattern formations and their active manipulation in Rydberg noisy-dressed Bose-Einstein condensates, {\it Opt. Lett.} {\bf 49}, 6517 (2024).

\bibitem{Hol22} S.~Hollerith {\it et al.}, Realizing distance-selective interactions in a Rydberg-dressed atom array, {\it Phys. Rev. Lett.} {\bf 128}, 113602 (2022).

\bibitem{Kun93} S.~Kunze {\it et al.}, Lifetime measurements of highly excited Rydberg states of strontium I, {\it Z. Phys. D} {\bf 27}, 111 (1993).

\bibitem{Kha24Terminal} M.~Khazali, Universal terminal for cloud quantum computing, {\it Sci. Rep.} {\bf 14}, 15412 (2024).

\bibitem{Kha20} M.~Khazali and K.~Mølmer, Fast multi-qubit gates via adiabatic evolution in dark state manifolds of Rydberg atoms and superconducting circuits, {\it Phys. Rev. X} {\bf 10}, 021054 (2020).

\bibitem{Kha25} M.~Khazali, Ultratight confinement of atoms in a Rydberg empowered optical lattice, {\it Quantum} {\bf 9}, 1585 (2025).

\bibitem{KhaFast24} M.~Khazali, Fast multicomponent cat-state generation under resonant or strong-dressing Rydberg-Kerr interaction, {\it Phys. Rev. A} {\bf 109}, 053716 (2024).

\bibitem{Saff}M.~Saffman, T. G. Walker, and K. M\o lmer. Quantum information with Rydberg atoms, {\it Reviews of modern physics} {\bf 82} 2313-2363 (2010).
\bibitem{Kha23Lech} M.~Khazali and W.~Lechner, Scalable quantum processors empowered by the Fermi scattering of Rydberg electrons, {\it Commun. Phys.} {\bf 6}, 57 (2023).

\bibitem{Kha23} M.~Khazali, All-optical quantum information processing via a single-step Rydberg blockade gate, {\it Opt. Express} {\bf 31}, 13970 (2023).

\bibitem{Mom25} A.~Momtaheni and M.~Khazali, Quantum computation with long-lived Rydberg-Landau atoms featuring suppressed ionization by the Magnetic Cage, arXiv:2506.00575 (2025).

\bibitem{Kha21IJAP} M.~Khazali, Quantum information and computation with Rydberg atoms, {\it Int. J. At. Phys.} {\bf 10}, 19 (2021).

\bibitem{Kha19} M.~Khazali, C.~Murry, and T.~Pohl, Polariton exchange interactions in multichannel optical networks, {\it Phys. Rev. Lett.} {\bf 123}, 113605 (2019).

\bibitem{Kha18} M.~Khazali, Progress towards macroscopic spin and mechanical superposition via Rydberg interaction, {\it Phys. Rev. A} {\bf 98}, 043836 (2018).

\bibitem{Kha17} M.~Khazali, K.~Heshami, and C.~Simon, Single-photon source based on Rydberg exciton blockade, {\it J. Phys. B: At. Mol. Opt. Phys.} {\bf 50}, 215301 (2017).

\bibitem{Kha16} M.~Khazali, H.~W.~Lau, A.~Humeniuk, and C.~Simon, Large Energy Superpositions via Rydberg Dressing, {\it Phys. Rev. A} {\bf 94}, 023408 (2016).

\bibitem{Kha15} M.~Khazali, K.~Heshami, and C.~Simon, Photon-photon gate via the interaction between two collective Rydberg excitations, {\it Phys. Rev. A} {\bf 91}, 030301(R) (2015).

\bibitem{KhaThesis}Khazali, M.  Applications of atomic ensembles for photonic quantum information processing and fundamental tests of quantum physics (Doctoral dissertation, University of Calgary (Canada)) (2016).

\end{thebibliography}
 \end{document}